\documentclass[]{spie}  

\usepackage{amsmath,amsfonts,amssymb}
\usepackage{graphicx}
\usepackage[colorlinks=true, allcolors=blue]{hyperref}
\usepackage{array}
\usepackage{makecell}

\DeclareMathOperator*{\argmax}{argmax}

\title{Axial multi-layer perceptron architecture for automatic segmentation of choroid plexus in multiple sclerosis}

\author[a,b]{Marius Schmidt-Mengin}
\author[a]{Vito A.G. Ricigliano}
\author[a,c]{Benedetta Bodini}
\author[a]{Emanuele Morena}
\author[a]{Annalisa Colombi}
\author[a]{Mariem Hamzaoui}
\author[a]{Arya Yazdan Panah}
\author[a,c]{Bruno Stankoff}
\author[a,b]{Olivier Colliot}

\affil[a]{Sorbonne Université, Paris Brain Institute, Inserm, CNRS, AP-HP, Paris, France}
\affil[b]{Inria, Aramis project-team, Paris, France}
\affil[c]{AP-HP, Hôpital Saint-Antoine, Department of Neurology, DMU Neurosciences, Paris, France}

\authorinfo{Further author information: (Send correspondence to O.C)\\O.C: E-mail: olivier.colliot@sorbonne-universite.fr}

\begin{document} 
\maketitle

\begin{abstract}
Choroid plexuses (CP) are structures of the brain ventricles which produce most of the cerebrospinal fluid (CSF). Several postmortem and in vivo studies have pointed towards their role in the inflammatory processes in multiple sclerosis (MS). Automatic segmentation of CP from MRI thus has high value for studying their characteristics in large cohorts of patients. To the best of our knowledge, the only freely available tool for CP segmentation is FreeSurfer but its accuracy for this specific structure is poor. In this paper, we propose to automatically segment CP from non-contrast enhanced T1-weighted MRI. To that end, we introduce a new model called "Axial-MLP" based on an assembly of Axial multi-layer perceptrons (MLPs).  This is inspired by recent works which showed that the self-attention layers of Transformers can be replaced with MLPs. This approach is systematically compared with a standard 3D U-Net, nnU-Net, Freesurfer and FastSurfer. For our experiments, we make use of a dataset of 141 subjects (44 controls and 97 patients with MS). We show that all the tested deep learning (DL) methods outperform FreeSurfer (Dice around 0.7 for DL vs 0.33 for FreeSurfer). Axial-MLP is competitive with U-Nets even though it is slightly less accurate. The conclusions of our paper are two-fold: 1) the studied deep learning methods could be useful tools to study CP in large cohorts of MS patients; 2)~Axial-MLP is a potentially viable alternative to convolutional neural networks for such tasks, although it could benefit from further improvements. An implementation is available at \url{https://github.com/aramis-lab/axial-mlp}.
\end{abstract}


\section{INTRODUCTION}
\label{sec:intro}  

Choroid plexuses (CP) are found in the brain ventricles and produce most of the cerebrospinal fluid (CSF). They are part of the blood–CSF barrier and a gateway for certain substances in and out of the brain. 
They have been shown to be associated with several diseases including Alzheimer's disease \cite{serot2003choroid, tadayon2020choroid} stroke \cite{egorova2019choroid},complex regional pain syndrome \cite{zhou2015enlargement} and leptomeningeal carcinomatosis \cite{mayinger2019mri}.
More specifically, there is a growing interest for the study of CP in multiple sclerosis (MS). Several post-mortem studies have shown the involvement of CP in the inflammatory process \cite{rodriguez2020inflammation,vercellino2008involvement,kooij2014disturbed}.
In vivo, signal changes in the CP have been reported in patients with MS \cite{kim2020choroid}. Very recently, it was demonstrated in vivo that CP are enlarged and inflamed in patients with MS \cite{ricigliano2021choroid}.
However, the role of CP in MS pathophysiology, as well as their potential relationship to disease evolution, are not entirely elucidated. There is thus a strong interest in studying CP in large populations of MS patients.

Manual segmentation of CP is time-consuming. As far as we know, the only freely available software for their automatic segmentation is FreeSurfer\cite{fischl2012freesurfer}. Nevertheless, its segmentation of CP is highly inaccurate \cite{tadayon2020improving, zhao2020choroid}. Hence, the development of more accurate automatic CP segmentation methods would be very useful to allow the study of large cohorts. 
To the best of our knowledge, only two other approaches have been proposed. The first one \cite{tadayon2020improving} clusters voxel intensities in order to separate CP voxels from CSF and ventricular wall inside the ventricles. The segmentation of the ventricles first needs to be obtained, for example by using FreeSurfer.
The second one \cite{zhao2020choroid} uses a 3D U-Net \cite{cciccek20163d, unet}. However, it included a very small sample of ten patients and only reported leave-one-out cross-validation results without evaluation on a proper independent testing set.
Moreover, neither of these two methods was designed or validated for patients with MS, so it is unclear how they would perform in this case.

Convolutional neural networks have been top performing in many computer vision benchmarks for several years. But recently, Transformer-based architectures \cite{vaswani2017attention} have been able to outperform convolutional models on several tasks.
Most recent approaches split an image into a grid of patches \cite{vit}, which are flattened to vectors 
and passed to a Transformer as a sequence for classification. SwinTransformers\cite{swin} enhanced this architecture for  segmentation and object detection. Some works have successfully applied such architectures to medical images \cite{unetr, transunet, cotr}. Even more recently, it has been shown that the self-attention layers of Transformers can be replaced by simple fully connected layers or multi-layer perceptrons (MLPs) \cite{melas2021you, resmlp, mlpmixer}. 

In this paper, we introduce a new model called "Axial-MLP", which is based on an assembly of Axial MLPs, for CP segmentation. We systematically compare this new approach to 3D U-Net, nnU-Net and Freesurfer.

\section{Materials and methods}

\subsection{Materials}

Data originated from three prospective studies performed between May 2009 and September 2017 at the Paris Brain Institute. The studies were approved by the local ethics committees and written informed consent was obtained from all participants. We studied 141 subjects: 44 healthy controls, 61 patients with relapsing remitting MS and 36 patients with progressive MS (13 patients with primary progressive MS and 23 patients with secondary progressive MS). 

Images were acquired\cite{ricigliano2021choroid} on two different Siemens 3T MRI scanners (Trio and Prisma) with a 32-channel head coil (92 participants on a Siemens Trio and 49 on a Siemens Prisma). 
The sequence used for this work was a 3D T1-weighted magnetization-prepared rapid gradient-echo
imaging (repetition time~=~2300ms, echo time~=~2.98ms). Other imaging sequences were acquired but were not used in this work. Dataset characteristics are reported in the first line of Table \ref{tab:dataset}. 

On each T1-weighted MRI, the choroid plexuses in the two lateral ventricles were segmented by a junior neurologist and corrected by a senior neurologist with 12 years of expertise in MRI processing. More details about image acquisition and manual segmentation procedures can be found in \cite{ricigliano2021choroid}. We divided the population using a stratified split into a training set (91 subjects) and a testing set (50 subjects). The training set was used for cross-validation in order to design and select the models. The testing set was left untouched until the very end of the study.

\begin{table}[ht]
\caption{Dataset characteristics: whole dataset, training set (used for cross-validation) and testing set.
F:~Female, M:~Male, T:~Siemens Trio MRI scanner, P:~Siemens Prisma MRI scanner, HC:~healthy control, R:~relapsing remitting MS, P1:~primary progressive MS, P2:~secondary progressive MS.
} 
\label{tab:dataset}
\begin{center}       
\begin{tabular}{lllll} 
\Xhline{3\arrayrulewidth}
\rule[-1ex]{0pt}{3.5ex}   & \multicolumn{1}{c}{Sex} & \multicolumn{1}{c}{Age} &

\multicolumn{1}{c}{Scanner} & \multicolumn{1}{c}{Disease Type}
\\
\Xhline{3\arrayrulewidth}
\rule[-1ex]{0pt}{3.5ex}  All & 141 (70F / 71M) & $40.7 \pm 12.5$ & 92T 49P & 44HC 61R 13P1 23P2
\\
\hline
\rule[-1ex]{0pt}{3.5ex}  Training & 91\hphantom{1} (45F / 46M) & $40.7 \pm 12.8$ & 60T 31P & 28HC 40R \hphantom{1}8P1 15P2 
\\
\hline
\rule[-1ex]{0pt}{3.5ex}  Testing & 50\hphantom{1} (25F / 25M) & $40.7 \pm 12.1$ & 32T 18P & 16HC 21R \hphantom{1}5P1 \hphantom{1}8P2 
\\
\Xhline{3\arrayrulewidth}
\end{tabular}
\end{center}
\end{table}

\subsection{Proposed Axial-MLP architecture for CP segmentation}
\label{sec:appendix}

Most Transformer-based vision models \cite{vit}  partition the input image into a grid of patches and then flatten these patches into embedding vectors. The resulting embeddings 
are then passed through a Transformer\cite{vaswani2017attention}. It has been shown that it is possible to replace the self-attention layers of the Transformer by simple fully-connected layers or MLPs with negligible loss of accuracy \cite{melas2021you, resmlp, mlpmixer}.
We posit that this may be particularly well-suited to our CP segmentation task, as we have registered all images to a common space, intuitively reducing the benefit of translation invariance. Furthermore, in order to reduce the number of parameters and to increase the inductive bias towards image processing, we apply our MLPs axially, as in Axial Attention Transformers \cite{axial}. We call the resulting architecture Axial-MLP.

\begin{figure} [ht]
   \begin{center}
   \begin{tabular}{c} 
   \includegraphics[width=1\textwidth]{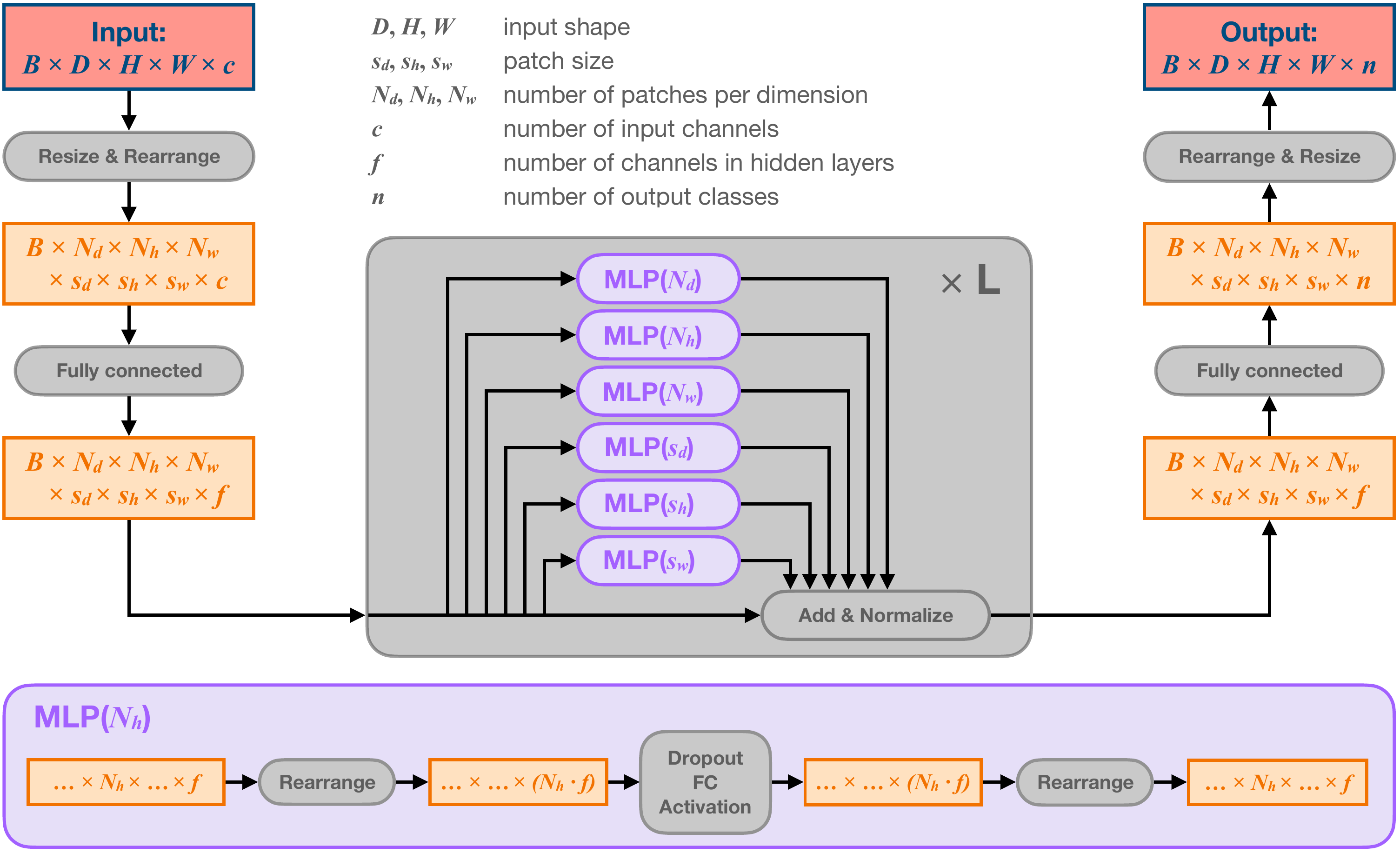}
   \end{tabular}
   \end{center}
   \caption[example] 
   { \label{fig:mlp} 
Schematic diagram of the proposed Axial-MLP for segmentation.}
   \end{figure}

Formally, we start by dividing our input image into non-overlapping patches, resulting in a tensor of shape
$B \times N_d \times N_h \times N_w \times s_d \times s_h \times s_w \times c$
where $B$ is the batch size, $N_i, i \in \{d, h, w\}$ are the number of patches along the depth, height and width, $s_i, \in \{d, h, w\}$ are the size of the patches along these dimensions, and $c$ is the number of channels of the input image. In practice, as the shape of the input image is not necessarily a multiple of the patch size, we start by downsampling it to a multiple of the patch size with trilinear interpolation.
Then, we expand this tensor along the channel axis with a fully connected layer. This is comparable to the first convolutional layer of a U-Net. We denote by $f$ the number of filters of this first layer. The resulting tensor has shape $B \times N_d \times N_h \times N_w \times s_d \times s_h \times s_w \times f$. We then pass this tensor through $L$ layers of our model. 
A layer consists of 6 fully connected layers. Each fully connected layer is applied along a specific dimension ($N_d, N_h, N_w, s_d, s_h$, or $s_w$) plus the channel dimension. The results are passed through a leaky ReLU (negative slope $= 0.01$) activation function and summed. The resulting tensor is normalized across all but the batch dimension (similarly to Layer Normalization \cite{layernorm}) and we apply a single trainable scalar weight and bias to the output (as opposed to Layer Normalization, where there is one such weight per element). Before each fully connected layer, we apply dropout with a drop rate of 0.02. The dropout is applied along the dimension of the corresponding fully connected layer, meaning that, similarly to Spatial Dropout\cite{spatialdrop}, the other dimensions are dropped together.
Finally, we rearrange the patches into an image and upsample the result so that is has the same shape as the original input image.
The whole model is depicted in Figure \ref{fig:mlp}.

If we assume a number of patches per dimension of $N$ and a patch size of $s$, then the number of parameters in the fully connected layers of Axial-MLP is proportional to $d(s^2+N^2)$, where $d$ is the number of dimensions of the input (3D in our case). Had we not applied the MLPs axially, as for example in ResMLP\cite{resmlp}, the number of parameters would be proportional to $s^{2d} + N^{2d}$. Thus, applying the MLPs axially allows to have a number of parameters that is linear instead of exponential in the number of dimensions, which makes a difference when going from 2D to 3D.

 We optimized this neural network with Adam \cite{kingma2014adam} ($\beta_1=0.9$, $\beta_2=0.999$, $\epsilon=10^{-8}$), with an initial learning rate of $10^{-2}$ decayed to $10^{-3}$ after 150 epochs. The training was stopped after 200 epochs. The batch size was 4. We used simple data  augmentations using affine transformations provided by TorchIO \cite{torchio} (scale $\in [0.9, 1.2]$, rotation $\in [-10, 10]$ degrees, translation $\in [-5, 5]$ voxels). In this paper, all Axial-MLPs have $L=6$ layers and the patch size is set to $8\times8\times8 $ voxels.

\subsection{Comparison to other approaches}
We compared our method to several approaches, ranging from trivial techniques to recent U-Net models. 

 \subsubsection{Mean and optimized mask}
 As the images are registered to a common space, we first established a trivial baseline that does not take into account the MRI images.
 Given a set of training ground truth masks $\mathcal{T}$, we computed
 \begin{equation*}
     \text{Mean mask} = \frac{1}{|\mathcal{T}|}\sum_{y\in\mathcal{T}} y \quad\quad
     \text{Optimized mask} \approx \argmax_{\overline{y}} \sum_{y\in\mathcal{T}} \text{Dice}(y, \overline{y})
\end{equation*}
 The optimized mask is computed approximately by starting from the mean mask and applying 100 gradient descent steps using the Adam optimizer \cite{kingma2014adam} (learning rate $1$, $\beta_1 = 0.9$, $\beta_2 = 0.999$, $\epsilon = 10^{-8}$).

\subsubsection{FreeSurfer and FastSurfer}
 We performed the segmentation using the $\texttt{recon-all}$ pipeline of FreeSurfer\cite{fischl2012freesurfer}. For comparison, we also applied FastSurfer, a deep learning-based approach that aims at reproducing the outputs of FreeSurfer while being considerably faster \cite{henschel2020fastsurfer}. Both of these methods were applied to the original, non registered, images.

\subsubsection{3D-UNet}
We trained a simple 3D U-Net \cite{cciccek20163d}. It's configuration was identical to the one described in the original paper, except that we varied the initial number of filters and used Instance Normalization \cite{instancenorm} and Leaky ReLU. We used the same optimizer and augmentations as for Axial-MLP, except that the learning rate was initially set to $10^{-3}$ instead of $10^{-2}$ and decayed to $10^{-4}$ instead of $10^{-3}$ after 150 epochs. The training was also stopped after 200 epochs and we used a batch size of 4.

 \subsubsection{nnU-Net}
 We also applied nnU-Net \cite{isensee2018nnu}, a method that aims to automatically determine many hyperparameters of a U-Net in order to optimize its performance on a given dataset. nnU-Net trains up to 3 different configurations: a 2D U-Net that operates on slices, a 3D U-Net that operates on the full resolution images but has to be trained on patches if the images are too large, and a 3D U-Net cascade in which the first U-Net operates on low resolution images, and the second one is applied at full resolution but patch-wise in order to refine the predictions of the first U-Net. In our case, the 3D U-Net cascade was not trained, as the images are small enough to be handled by the full resolution 3D U-Net without patches. For the 2D and 3D U-Net, the following hyperparameters were determined by nnU-Net: no resampling, 4 downsamplings/upsamplings, 32 features in the first stage (doubled in every stage), batch size of 4 for 3D and 308 for 2D. The 3D nnU-Net trainings failed to stop by themselves within 48 hours, so they were killed.

\subsection{Preprocessing}
For FreeSurfer and FastSurfer, we did not preprocess the images. For all other methods, we ran the $\texttt{t1-linear}$ pipeline of Clinica\cite{routier2021clinica,wen2020convolutional}. It performs bias field correction using the N4ITK method\cite{tustison2010n4itk} and affine registration using the SyN algorithm\cite{avants2008symmetric} from ANTs\cite{avants2014insight} to align each image to the MNI space with the ICBM 2009c nonlinear symmetric template\cite{fonov2009unbiased,fonov2011unbiased}.
The resulting images have 1 mm isotropic voxels.
We then applied the computed transformations to the segmentation masks using a nearest neighbor interpolation kernel. 
Finally, we cropped all images using a bounding box of $102\times 94 \times 76$ voxels that encompasses all segmentation masks from the training set plus a 10 voxel margin.
For all neural network trainings, we first apply a $z$-normalization to each image.

\clearpage
\subsection{Evaluation}

All experiments were performed using 32GB Nvidia V100 GPUs. 

\subsubsection{Evaluation metrics}

We reported the following metrics:
\begin{equation*}
    \text{Precision }P = \frac{\sum_{i\in I} x_i y_i}{\sum_{i\in I} x_i} \quad\quad
    \text{Recall }R = \frac{\sum_{i\in I} x_i y_i}{\sum_{i\in I} y_i} \quad\quad 
    \text{Dice score }D = \frac{2\sum_{i\in I} x_i y_i}{\sum_{i\in I} x_i + y_i} = \frac{2}{\frac{1}{P} + \frac{1}{R}}
\end{equation*}
\begin{equation*}
    \text{Volume error rate }\mathit{VER} = \frac{\sum_{i\in I} x_i  - y_i}{\sum_{i\in I} y_i} \quad\quad
    \text{Absolute volume error rate }\mathit{AVER} = |\mathit{VER}|
\end{equation*}
where we denote by $I$ the set of voxels of a given image, $x_i$ is the prediction for voxel $i\in I$ and $y_i$ the corresponding ground truth label.

We also reported Pearson's $r$ between the predicted volumes and the ground truth volumes. All these formulas remain valid and make sense when $x\in [0, 1]$. Hence, we chose not to binarize the output of our models, as we believe that it would unnecessarily destroy some information.

The volume error rate, once averaged over the dataset, reflects whether a particular method tends to over or under-estimate the volume, whereas the absolute volume error rate measures the overall accuracy of the estimated volumes.
It can be shown that $D-1 \leq \mathit{VER} \leq 1/D-1$, so that a high enough Dice score ensures a low volume error rate.

 \subsubsection{Validation procedure}
 
 We first used a 5-fold cross-validation (CV) on the training set (except for FreeSurfer and FastSurfer, for which there is no need to use CV as they do not require any form of tuning). This means that the training set was split into training and validation sets while the test set was left untouched.
We used the same 5 splits for all methods.
On each fold, the model with best validation accuracy over the course of the training was selected to perform inference on the fold's validation set, and the predictions of all 5 folds were aggregated into one large validation set, over which the evaluation metrics were computed.
Using the best model of each fold can cause the CV metrics to be over-estimated, but this is not problematic as we have a testing set that was not used until the end of the study and only used for the final evaluation. For inference on this testing set, we averaged the predictions from the best models of each CV fold.

\section{Results}

Cross-validation results are presented in Table~\ref{tab:CVresults}.
We experimented with different configurations of the standard 3D U-Net (4, 8, 16 and 32 initial filters) and the proposed Axial-MLP (4, 8 and 16 hidden channels).
Their cross-validation Dice coefficients, as well as resource consumption, are reported in Table~\ref{tab:mlp_unet_results}.
We judged that Axial-MLP 8 and 3D U-Net 8 offered the best CV accuracy vs training time compromise, and therefore only these architectures were applied to the testing set (in order not to overfit this hyperparameter). 3D U-Net models were slightly better than Axial-MLP in terms of Dice coefficient, but their number of parameters is also higher.

\begin{table}[h]
 \caption
   { \label{tab:CVresults} 
5-fold cross-validation results on the training set.  
Note that the reported standard deviations are the empirical values computed over the 5 folds and are thus biased estimates of the standard-deviation. 
Pearson's $r$ is not applicable to the mean and optimized mask methods as these are constant across all samples.
MVER:~Mean volume error rate. MAVER:~Mean absolute volume error rate.}
\begin{center}
\begin{tabular}{lcccccc} 
\Xhline{3\arrayrulewidth}
\rule[-1ex]{0pt}{3.5ex} & Dice & Precision & Recall & MVER & MAVER & \begin{tabular}[x]{@{}c@{}} Pearson's $r$\end{tabular}\\ 
\Xhline{3\arrayrulewidth}
\rule[-1ex]{0pt}{3.5ex}Mean mask & $0.22 \pm 0.05 $ & $0.22 \pm 0.07 $ & $0.23 \pm 0.05 $ & $0.10 \pm 0.36 $ & $0.26 \pm 0.26 $ & N/A \\ \hline
\rule[-1ex]{0pt}{3.5ex}Optimized mask & $0.38 \pm 0.11 $ & $0.29 \pm 0.10 $ & $0.57 \pm 0.16 $ & $1.08 \pm 0.68 $ & $1.09 \pm 0.68 $ & N/A \\
\Xhline{3\arrayrulewidth}
\rule[-1ex]{0pt}{3.5ex}FreeSurfer & $0.33 \pm 0.09 $ & $0.42 \pm 0.11 $ & $0.28 \pm 0.09 $ & $-0.32 \pm 0.22 $ & $0.35 \pm 0.17 $ & $0.65 $\\ \hline
\rule[-1ex]{0pt}{3.5ex}FastSurfer & $0.34 \pm 0.09 $ & $0.41 \pm 0.11 $ & $0.30 \pm 0.09 $ & $-0.24 \pm 0.24 $ & $0.29 \pm 0.17 $ & $0.57 $\\
\Xhline{3\arrayrulewidth}
\rule[-1ex]{0pt}{3.5ex}nnU-Net 2D & $0.70 \pm 0.06 $ & $0.69 \pm 0.09 $ & $0.72 \pm 0.07 $ & $0.08 \pm 0.26 $ & $0.17 \pm 0.22 $ & $0.77 $\\ \hline
\rule[-1ex]{0pt}{3.5ex}nnU-Net 3D & $0.73 \pm 0.06 $ & $0.72 \pm 0.09 $ & $0.75 \pm 0.06 $ & $0.08 \pm 0.26 $ & $0.16 \pm 0.23 $ & $0.82 $\\ \hline
\rule[-1ex]{0pt}{3.5ex}3D U-Net 8 & $0.72 \pm 0.06 $ & $0.69 \pm 0.10 $ & $0.77 \pm 0.06 $ & $0.17 \pm 0.34 $ & $0.20 \pm 0.32 $ & $0.77 $\\ \hline
\rule[-1ex]{0pt}{3.5ex}Axial-MLP 8 (proposed) & $0.71 \pm 0.06 $ & $0.67 \pm 0.10 $ & $0.78 \pm 0.06 $ & $0.21 \pm 0.35 $ & $0.24 \pm 0.33 $ & $0.75 $\\ 
\Xhline{3\arrayrulewidth}
\rule[-1ex]{0pt}{3.5ex}
\end{tabular}
\end{center}
\end{table}

\begin{table}[h]
 \caption
   { \label{tab:test_results} 
Results on the independent testing set. Pearson's $r$ is not applicable to the mean and optimized mask methods as these are constant across all samples.
MVER:~Mean volume error rate. MAVER:~Mean absolute volume error rate.}
\vspace{-4mm}
\begin{center}
\begin{tabular}{lcccccc} 
\Xhline{3\arrayrulewidth}
\rule[-1ex]{0pt}{3.5ex}& Dice & Precision & Recall & MVER & MAVER & \begin{tabular}[x]{@{}c@{}} Pearson's $r$\end{tabular}\\
\Xhline{3\arrayrulewidth}
\rule[-1ex]{0pt}{3.5ex}Mean mask & $0.23 \pm 0.05 $ & $0.24 \pm 0.07 $ & $0.24 \pm 0.04 $ & $0.13 \pm 0.43 $ & $0.29 \pm 0.34 $ & N/A \\ \hline
\rule[-1ex]{0pt}{3.5ex}Optimized mask & $0.41 \pm 0.10 $ & $0.31 \pm 0.10 $ & $0.62 \pm 0.14 $ & $1.16 \pm 0.82 $ & $1.16 \pm 0.82 $ & N/A \\
\Xhline{3\arrayrulewidth}
\rule[-1ex]{0pt}{3.5ex}FreeSurfer & $0.33 \pm 0.08 $ & $0.42 \pm 0.10 $ & $0.29 \pm 0.08 $ & $-0.29 \pm 0.18 $ & $0.32 \pm 0.14 $ & $0.65 $\\ \hline
\rule[-1ex]{0pt}{3.5ex}FastSurfer & $0.35 \pm 0.09 $ & $0.40 \pm 0.10 $ & $0.32 \pm 0.10 $ & $-0.18 \pm 0.23 $ & $0.25 \pm 0.15 $ & $0.59 $\\ 
\Xhline{3\arrayrulewidth}
\rule[-1ex]{0pt}{3.5ex}nnU-Net 2D & $0.70 \pm 0.06 $ & $0.68 \pm 0.10 $ & $0.73 \pm 0.07 $ & $0.11 \pm 0.27 $ & $0.20 \pm 0.21 $ & $0.74 $\\ \hline
\rule[-1ex]{0pt}{3.5ex}nnU-Net 3D & $\mathbf{0.72} \pm 0.05 $ & $\mathbf{0.72} \pm 0.10 $ & $0.75 \pm 0.06 $ & $\mathbf{0.09} \pm 0.26 $ & $\mathbf{0.20} \pm 0.19 $ & $0.79 $\\ \hline
\rule[-1ex]{0pt}{3.5ex}3D U-Net 8 & $0.72 \pm 0.06 $ & $0.68 \pm 0.10 $ & $0.78 \pm 0.06 $ & $0.19 \pm 0.31 $ & $0.23 \pm 0.28 $ & $\mathbf{0.80} $\\ \hline
\rule[-1ex]{0pt}{3.5ex}Axial-MLP 8 (proposed) & $0.71 \pm 0.06 $ & $0.65 \pm 0.10 $ & $\mathbf{0.79} \pm 0.06 $ & $0.25 \pm 0.32 $ & $0.29 \pm 0.29 $ & $0.76 $\\
\Xhline{3\arrayrulewidth}
\rule[-1ex]{0pt}{4ex}
\end{tabular}
\end{center}
\end{table}

\begin{table}[h!]
 \caption
   { \label{tab:mlp_unet_results} Different characteristics of 3D U-Net and Axial-MLP when varying the number of initial filters/hidden channels. The Dice coefficient is from the CV. The reported times represent the duration of a training step with a batch size of 4 in seconds. The GPU memory consumption, given in MB, was obtained with the \texttt{nvidia-smi} command and a batch size of 4. We also tried to train a 32 channel Axial-MLP but ran out of GPU memory on a 32GB Nvidia V100.}
\vspace{-4mm}
\begin{center}
\setlength{\tabcolsep}{0.3em}
\begin{tabular}[t]{lcccc} 
\Xhline{3\arrayrulewidth}
\rule[-1ex]{0pt}{3.5ex}& Dice & Params & Time & Memory\\
\Xhline{3\arrayrulewidth}
\rule[-1ex]{0pt}{3.5ex}U-Net 4 & $ 0.711 \pm 0.060 $ & 256,813 & 0.179 & 3,930 \\ \hline
\rule[-1ex]{0pt}{3.5ex}U-Net 8 & $ 0.719 \pm 0.060 $ & 1,021,505 & 0.282 & 6,418 \\ \hline
\rule[-1ex]{0pt}{3.5ex}U-Net 16 & $ 0.722 \pm 0.061 $ & 4,082,049 & 0.596 & 11,026 \\ \hline
\rule[-1ex]{0pt}{3.5ex}U-Net 32 & $ 0.723 \pm 0.058 $ & 16,320,257 & 1.682 & 20,668 \\
\Xhline{3\arrayrulewidth}
\rule[-1ex]{0pt}{3.5ex}Axial-MLP 4 \quad & $ 0.704 \pm 0.062 $ & 53,017 & 0.114 & 5,026 \\ \hline
\rule[-1ex]{0pt}{3.5ex}Axial-MLP 8 \quad & $ 0.711 \pm 0.060 $ & 209,317 & 0.229 & 8,614 \\ \hline
\rule[-1ex]{0pt}{3.5ex}Axial-MLP 16 \quad & $ 0.713 \pm 0.058 $ & 831,805 & 0.474 & 15,654 \\
\Xhline{3\arrayrulewidth}
\end{tabular}
\end{center}
\end{table}

\begin{figure} [h]
   \begin{center}
   \begin{tabular}{c} 
   \includegraphics[width=1\textwidth]{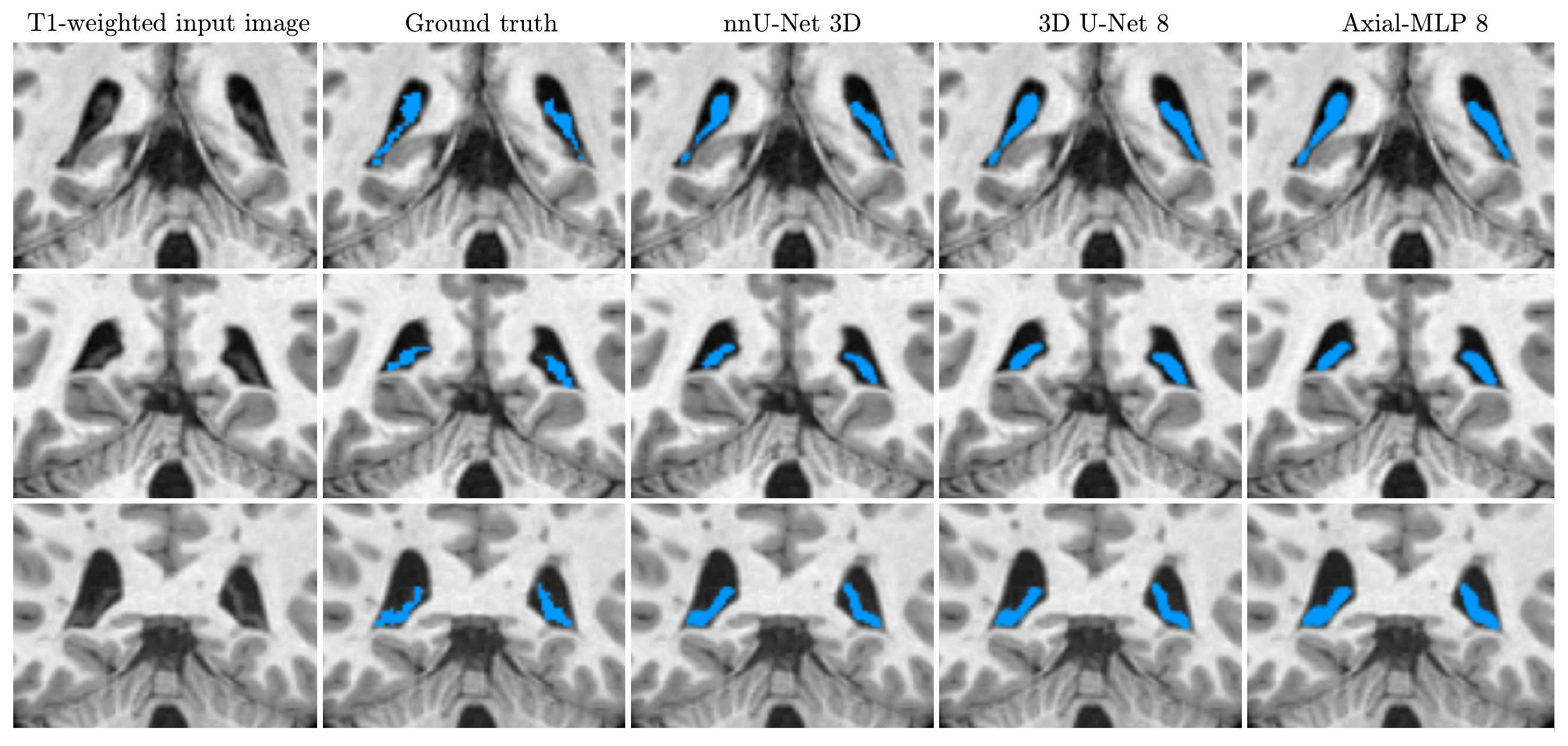}
   \end{tabular}
   \end{center}
   \vspace{-5mm}
   \caption[example] 
   { \label{fig:nn} 
Examples of nnU-Net, 3D U-Net and Axial-MLP segmentation in coronal view for three patients.}
\end{figure}

\begin{figure} [h!]
\vspace{4mm}
   \begin{center}
   \begin{tabular}{c} 
   \includegraphics[width=1\textwidth]{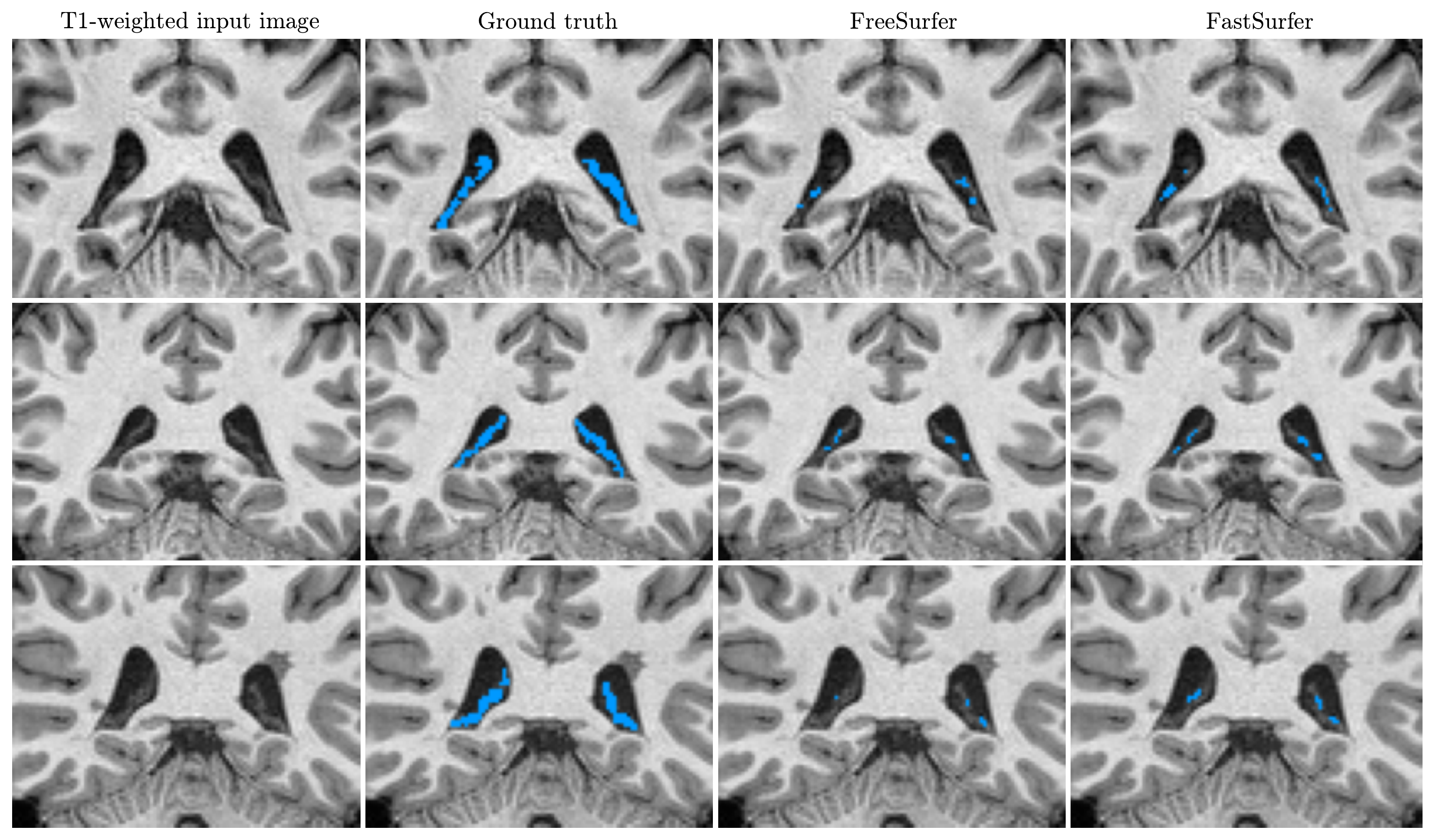}
   \end{tabular}
   \end{center}
   \vspace{-5mm}
   \caption[example] 
   { \label{fig:fs} 
Examples of FreeSurfer and FastSurfer segmentation in coronal view for three patients.}
   \end{figure}

Results on the testing set are presented in Table~\ref{tab:test_results}.
As expected, FreeSurfer and FastSurfer resulted in poor segmentation quality (Dice around 0.3-0.4).
On the testing set, all deep learning (DL) models outperformed FreeSurfer by a large margin (Dice between 0.71 and 0.72, vs 0.33-0.35 for FreeSurfer and FastSurfer). Dice  coefficients of all DL models were comparable. Nevertheless, nnU-Net performed slightly better on almost all metrics. The computation time it used for training was considerably higher (nnU-Net was trained for 48(3D)+12(2D) hours per fold, instead of under 2 hours per fold for all U-Nets and Axial-MLPs). The difference in Dice score between 3D U-Net 8 and Axial-MLP 8 (respectively between nnU-Net 3D and Axial-MLP 8) was small but statistically significant (3D U-Net vs Axial-MLP 8, Dice: 0.72 vs 0.71, paired Student's t-test, $T=4.8$, $p<10^{-4}$; nnU-Net 3D vs Axial-MLP 8, Dice: 0.72 vs 0.71, paired Student's t-test, $T=6.5$, $p<10^{-7}$).
Anecdotally, the constant solution "optimized mask" had about the same Dice as Freesurfer. 
Illustrative examples of automatic and manual CP segmentations are presented on Figure~\ref{fig:fs} for FreeSurfer and FastSurfer and on Figure~\ref{fig:nn} for the various deep learning models.

\section{Discussion and conclusion}

In this paper, we proposed a new architecture called "Axial-MLP" for segmentation of choroid plexuses on MRI. It was competitive with state-of-the-art segmentation methods and achieved results which were comparable to those of 3D U-Nets and nnU-Net. Its performance, as assessed by the Dice coefficient, was nevertheless slightly lower (0.71 vs 0.72). Overall, our work demonstrates that MLP models can be competitive with convolutional neural networks for medical image segmentation.
We believe that this is useful information for the community and that future improved MLP-based architectures have the potential to outperform convolutional models.

To the best of our knowledge, the only freely available tool for automatic CP segmentation is Freesurfer. Consistently with previous studies, we showed that it is inaccurate for this specific brain structure \cite{tadayon2020improving, zhao2020choroid}. As expected, FastSurfer required much less computational time than FreeSurfer.
It was also slightly more accurate.
All deep learning models vastly outperformed FreeSurfer. Interestingly, in our specific application, nnU-Net did not substantially improve the main performance metrics compared to a standard 3D U-Net even though it required considerably more computational time. 

The trained deep learning models resulted in good overlap and volume agreement with the manual reference. They could potentially be useful tools to study choroid plexuses in large cohorts of patients with multiple sclerosis in order to shed further light on their role in disease evolution. Nevertheless, this will require further validation on larger datasets with more variable image acquisition settings.

\acknowledgments 
The research leading to these results has received funding from the French government under management of Agence Nationale de la Recherche as part of the ``Investissements d'avenir'' program, reference ANR-19-P3IA-0001 (PRAIRIE 3IA Institute) and reference ANR-10-IAIHU-06 (Agence Nationale de la Recherche-10-IA Institut Hospitalo-Universitaire-6).  This work was granted access to the HPC resources of IDRIS (Jean Zay supercomputer) under the allocation "Accès dynamique" 102000 (number: AD011012863) made
by GENCI (Grand Équipement National de Calcul Intensif). The authors are grateful to Mauricio Diaz-Melo for technical assistance to access the Jean Zay supercomputer.

\bibliography{report} 
\bibliographystyle{spiebib} 

\end{document}